\documentclass[aps, prd, twocolumn, nofootinbib, showpacs]{revtex4}
\pdfoutput=1
\usepackage{amsfonts,amsmath,amsthm,amssymb,graphicx,hyperref}

\begin{document}

\preprint{1234}

\title{Tachyacoustic Cosmology: An Alternative to Inflation}

\author{Dennis Bessada${}^{1,2}$,\footnote{
        {\tt dbessada@buffalo.edu},\\
        ${}^\dagger\,{}${\tt whkinney@buffalo.edu},\\${}^\ddagger\,{}${\tt ds77@buffalo.edu},\\${}^\S\,{}${\tt jwang@niagara.edu}}
        William~H.~Kinney${}^{1,\dagger}$, Dejan Stojkovic ${}^{1,\ddagger}$ and John Wang${}^{1,3\S}$}

     \affiliation{
              ${}^1$Dept. of Physics, University at Buffalo, the  State University of New York, Buffalo, NY 14260-1500, United States\\
              ${}^2$INPE - Instituto Nacional de Pesquisas Espaciais - Divis\~ao de Astrof\'isica, S\~ao Jos\'e dos Campos, 12227-010 SP, Brazil\\
              ${}^3$Dept. of Physics, Niagara University, NY 14109-2044, United States}

\begin{abstract}

We consider an alternative to inflation for the generation of
superhorizon perturbations in the universe in which the speed of
sound is faster than the speed of light. We label such cosmologies,
first proposed by Armendariz-Picon, {\it tachyacoustic}, and explicitly
construct examples of non-canonical Lagrangians which have
superluminal sound speed, but which are causally self-consistent.
Such models possess two horizons, a Hubble horizon and an acoustic
horizon, which have independent dynamics. Even in a decelerating
(non-inflationary) background, a nearly scale-invariant spectrum
of perturbations can be generated by quantum perturbations
redshifted outside of a shrinking acoustic horizon. The acoustic
horizon can be large or even infinite at early times, solving the
cosmological horizon problem without inflation. These models do
not, however, dynamically solve the cosmological flatness problem,
which must be imposed as a boundary condition. Gravitational wave
modes, which are produced by quantum fluctuations exiting the
Hubble horizon, are not produced.

\end{abstract}


\maketitle

\section{Introduction: Why Inflation Works}
\label{sec:introduction}

Inflationary cosmology \cite{Guth:1980zm,Linde:1981mu,Albrecht:1982wi} is the most successful and widely accepted paradigm for understanding the very
early universe. By all measures inflation is a compelling and scientifically useful theory, and makes quantitative predictions which have been strongly
supported by recent data \cite{Dunkley:2008ie,Komatsu:2008hk,Kinney:2008wy}. The two main hallmarks of inflationary cosmology are solutions to the
flatness and horizon problems of the standard Big Bang cosmology: why is the universe so close to geometrically flat, and how did the apparent acausal
structure of the universe arise? ``Acausal'' more specifically means that the universe is approximately homogeneous on scales larger than a Hubble
length $H^{-1}$, and in addition exhibits a spectrum of density perturbations which is correlated on scales larger than a Hubble length.
Such {\it superhorizon} correlations are generated in inflation by accelerated expansion, $\ddot a / a > 0$, where $a\left(t\right)$ is the
cosmological scale factor, which means that the comoving Hubble length $d_H \simeq (a H)^{-1}$ shrinks with the expansion of the universe,
\begin{equation}
\frac{d}{d \ln{a}} \left(a H\right)^{-1} < 0.
\end{equation}
Therefore, quantum perturbations, which have constant wavelength in comoving units, are smaller than the Hubble length at early times, and are
redshifted to larger than the Hubble length at late times, where they are ``frozen'' as classical perturbations. Furthermore, as long as the Hubble
constant $H$ is slowly varying with time, the perturbations generated in inflation are nearly scale-invariant, consistent with observation.
Furthermore, the solution to the horizon problem and the flatness problem are linked in inflation via a conservation law,
\begin{equation}
\label{eq:conservationlaw}
\frac{d}{d \ln{a}} \frac{\left\vert\Omega - 1\right\vert}{d_H^2} = 0.
\end{equation}
Through this conservation law, a universe with shrinking comoving horizon size is identical to a universe which is evolving toward flatness,
\begin{equation}
\frac{d}{d \ln{a}} \left\vert\Omega - 1\right\vert < 0.
\end{equation}
Inflation therefore solves the horizon and flatness problems of the standard Big Bang with a {\it single} mechanism: accelerated expansion.

However, inflation is not the only way to accomplish this goal, as
can be seen from the fact that the acceleration $\ddot a$ appears
nowhere in the conservation law (\ref{eq:conservationlaw}). To
solve both the horizon and flatness problems, it is sufficient to
have a shrinking comoving Hubble length. One way to do this is
accelerated expansion, but another is to have a collapsing
universe, $H = \left(\dot a / a\right) < 0$. A collapsing, matter-
or radiation-dominated universe also has a shrinking comoving
Hubble length, which will generate perturbations in a manner
similar to inflation. This is the mechanism used by the Ekpyrotic
scenario \cite{Gratton:2003pe} to construct a cosmology consistent
with observations. It is also possible to decouple the horizon and
flatness problems, for example in theories with a varying speed of
light, so that the causal horizon is much larger than the Hubble
length \cite{Albrecht:1998ir}. Such theories can in principle
solve the horizon problem, but not the flatness problem, since the
conservation law (\ref{eq:conservationlaw}) is violated. It is
also possible to solve the horizon problem by a universe which is
much older than a Hubble time as in string gas cosmology
\cite{Brandenberger:2009jq} or island cosmology \cite{Dutta:2005gt,Dutta:2005if}, or by the inclusion of extra
dimensions \cite{Starkman:2000dy,Starkman:2001xu}. However, it has been argued that
inflation and ekpyrosis are the {\it only} mechanisms for
generating a scale-invariant spectrum of perturbations
\cite{Gratton:2003pe,Khoury:2008wj}.

In this paper, we discuss a method of solving the cosmological horizon problem and seeding scale-invariant primordial perturbations in a cosmology with decelerating expansion and a corresponding {\it growing} comoving Hubble horizon. The key to implementing such a model is the fact that curvature perturbations are not generated at the Hubble horizon, but at the acoustic horizon determined by the speed of sound of a scalar field. For canonical field theories, the two are identical, but for non-canonical field theories, they are not. If one has a decaying, superluminal sound speed, curvature perturbations can be generated outside the Hubble horizon without inflation.  We propose the term
 {\it tachyacoustic} for such cosmologies, which are closely related to varying speed of light theories. This idea has some history: such cosmologies were first proposed by Armendariz-Picon in the context of modified dispersion relations \cite{ArmendarizPicon:2006if}, and the generation of perturbations in such cosmologies was further considered by Piao \cite{Piao:2006ja}. The idea re-emerged in the context of varying speed of light theories by Magueijo \cite{Magueijo:2008pm}, and non-canonical Lagrangians by Magueijo \cite{Magueijo:2008sx} and Piao \cite{Piao:2008ip}. In this paper, we outline a general approach to such cosmologies based on the generalization of the inflationary flow formalism \cite{Kinney:2002qn} introduced by Bean, {\it et al.} for the case of arbitrary Lagrangians \cite{Bean:2008ga}. We find that there is a class of Lagrangians with the necessary properties for tachyacoustic cosmologies, and discuss two interesting examples. We find that it is straightforward to generate nearly scale-invariant perturbations for these Lagrangians, and show that they have the property of reducing to instanton-like solutions with infinite sound speed on the initial-time boundary of the spacetime. We speculate that this property may allow a self-consistent description of tachyacoustic cosmologies within a Wheeler-deWitt description of quantum cosmology. Finally, we show that such models are causally self-consistent, and argue that they form a viable class of alternatives to inflation.

\section{Tachyacoustic Cosmology}
\label{sec:TachyacousticCosmology}

In this paper, we consider a  way of generating scale-invariant superhorizon cosmological perturbations based on non-canonical scalar field
Lagrangians with a speed of sound faster than the speed of light, $c_S > 1$.  If the universe is dominated by a scalar field with speed of sound $c_S$, the relevant horizon for
the generation of density perturbations is not the Hubble horizon $d_H \simeq (a H)^{-1}$ but the acoustic horizon,
\begin{equation}
D_H \simeq \frac{c_S}{a H}.
\end{equation}
Mode freezing at the acoustic horizon is well-known in
non-canonical inflation models, for example k-Inflation
\cite{ArmendarizPicon:1999rj} and DBI inflation
\cite{Silverstein:2003hf}. In non-canonical inflation models, the
Hubble horizon and the acoustic horizon are {\it both} shrinking
in comoving units, resulting in the generation of density
perturbations at the acoustic horizon and gravitational wave
perturbations at the Hubble horizon \cite{Garriga:1999vw}.
However, the comoving Hubble horizon need not be shrinking to
generate curvature perturbations: all that is required is that the
{\it acoustic} horizon be shrinking, $d D_H / d \ln{a} < 0$. In
this case, if curvature perturbations are to be generated on
scales larger than the Hubble horizon, it is necessary that the
acoustic horizon be larger than the Hubble horizon, which requires
a speed of sound greater than the speed of light. Such theories
were studied recently by Babichev {\it et al.}
\cite{Babichev:2006vx,Babichev:2007dw}, who showed that k-essence theories with
$c_S > 1$ are causally self-consistent (see Appendix
\ref{sec:stablecau}), and can be mapped to bimetric theories with
two ``light cones'', one given by the Hubble horizon, and the
other given by the acoustic horizon, which can be larger than the
Hubble horizon without the presence of closed timelike loops. This
opens the possibility that one can construct a decelerating
cosmology which nonetheless generates perturbations on
super-Hubble scales via a superluminal, shrinking acoustic cone.

To explicitly construct such a model, consider a Dirac-Born-Infeld (DBI) Lagrangian,
\begin{equation}
\label{eq:DBIlagrangian}
{\cal L}\left(\phi,X\right) = - f^{-1}\left(\phi\right) \sqrt{1 - f\left(\phi\right) \dot\phi^2} + f^{-1}\left(\phi\right) - V\left(\phi\right),
\end{equation}
where we take the metric to have negative signature $\eta_{\mu\nu} = {\rm diag.}(1, -1, -1, -1)$, so that the kinetic term for the field is positive,
\begin{equation}
\dot\phi^2 \equiv g^{\mu\nu} \partial_\mu \phi \partial_\nu \phi > 0.
\end{equation}
Such Lagrangians often arise in string theory, for example in the case of DBI inflation. For the moment, we will not attempt to make a connection with
string theory, but will take the Lagrangian (\ref{eq:DBIlagrangian}) as a phenomenological {\it ansatz}. Take the background spacetime to be a flat,
Friedmann-Robertson-Walker (FRW) space, $g_{\mu \nu} = a^2 \eta_{\mu\nu}$, so that the scale factor evolves as
\begin{equation}
a \propto \exp\left(\int{H dt}\right) \propto e^{- N},
\end{equation}
where we define the number of e-folds $N$ as\footnote{We use the usual convention that $N \rightarrow \infty$ corresponds to early time, and
$N \rightarrow - \infty$ corresponds to late time.}
\begin{equation}
N \equiv - \int{H dt}.
\end{equation}
There is a class of exact solutions \cite{Chimento:2007es,Kinney:2007ag} to the equation of motion for the field $\phi$ characterized by two
dimensionless flow parameters $\epsilon$ and $s$, where
\begin{equation}
\label{eq:defofepsilon}
\epsilon \equiv \frac{1}{H} \frac{d H}{d N} = {\rm const.},
\end{equation}
and
\begin{equation}
s \equiv -\frac{1}{c_S}\frac{d c_S}{d N} = {\rm const.}
\end{equation}
The parameter $\epsilon$ has its usual interpretation in terms of the equation of state of the scalar field,
\begin{equation}
\label{eq:epsilondef} p = \rho \left(\frac{2}{3} \epsilon -
1\right).
\end{equation}
For $\epsilon = {\rm const.}$, the scale factor evolves as a power-law, $a \propto t^{1 / \epsilon}$, so that the expansion is accelerating
({\it i.e.} inflation) for $\epsilon < 1$. The speed of sound evolves as
\begin{equation}
\label{eq:DBIsoundspeed}
c_S = \sqrt{1 - f\left(\phi\right) \dot\phi^2} \propto e^{-s N},
\end{equation}
and the Hubble parameter evolves as
\begin{equation}
\label{eq:Hphi} H = \frac{\dot a}{a} \propto e^{\epsilon N}.
\end{equation}
The parameter $\epsilon$ is a positive-definite quantity for $p \geq -\rho$, so that the Hubble constant always decreases with expansion. In contrast,
the parameter $s$ can take either sign, with $s > 0$ corresponding to a sound speed which increases with expansion, and $s < 0$ corresponding to an
decreasing sound speed. (See Ref. \cite{Kinney:2007ag} for a detailed derivation of this solution.) The important dynamics for the generation of
perturbations is the time evolution of the corresponding horizons in comoving units. The comoving Hubble horizon evolves as
\begin{equation}
d_H \propto \left(a H\right)^{-1} \propto e^{\left(1 - \epsilon\right) N} \propto \tau,
\end{equation}
where $\tau$ is the conformal time. The Hubble horizon is shrinking in comoving units for $\epsilon < 1$, which is identical to accelerated expansion,
and is the usual condition for inflation. The acoustic horizon behaves as
\begin{equation}
D_H \propto \frac{c_S}{a H} \propto e^{\left(1 - \epsilon - s\right) N} \propto \tau^{\left(1 - \epsilon - s\right) / \left(1 - \epsilon\right)}.
\end{equation}
Therefore the condition for a shrinking acoustic horizon, $1 -
\epsilon - s > 0$, is {\it not} identical to accelerated
expansion. For $\epsilon > 1$ and $s < 1 - \epsilon$, the
expansion is non-inflationary, the Hubble horizon is growing in
comoving units, and the acoustic horizon is shrinking. The initial
singularity is at $\tau = 0$, and we see immediately that for the
tachyacoustic solution, the speed of sound in the scalar field is
{\it infinite} at the initial singularity, and the acoustic
horizon is likewise infinite in size. Therefore, such a cosmology
presents no ``horizon problem'' in the usual sense, since even a
spatially infinite spacetime is causally connected on the
initial-time boundary. Furthermore, unlike in the case of
inflation, there is no period of reheating necessary, since the
cosmological evolution can be radiation-dominated throughout and
the cosmic temperature is not driven exponentially to zero.

In the next section, we use the the generalized flow function approach of Bean, {\it et al.} \cite{Bean:2008ga} to construct a class of Lagrangians
with solutions of the type outlined above, with constant flow parameters. In these solutions, the scale-factor evolves as a power-law in time and the
equation of motion for curvature perturbations can be solved exactly, which we discuss in Sec. \ref{sec:Perturbations}.

\section{Flow hierarchy for general k-essence models}
\label{sec:flowhier}

We now generalize the discussion in the last section to an arbitrary k-essence model. Consider a general Lagrangian of the form
${\cal{L}}={\cal{L}}\left[X,\phi\right]$, where $2X=g^{\mu\nu}\partial_{\mu}\phi\partial_{\nu}\phi$ is the canonical kinetic term ($X>0$ according
to our choice of the metric signature). The energy density $\rho$ and pressure $p$ are given by
\begin{eqnarray}
p &=& {\cal L}\left(X,\phi\right),\\
\rho &=& 2 X {\cal L}_X - {\cal L}.
\end{eqnarray}
The speed of sound is given by
\begin{eqnarray}
\label{defspeedofsound} c_S^{2} &\equiv& \frac{p_X}{\rho_X} = \left(1 + 2X\frac{{\cal
L}_{XX}}{{\cal L}_{X}}\right)^{-1},
\end{eqnarray}
where the subscript ``${X}$" indicates a derivative with respect to the kinetic
term.  Throughout this section, unless otherwise stated, we will follow closely Bean {\it{et al.}} \cite{Bean:2008ga}. We define the first three
{\it flow parameters} as derivatives with respect to the number of e-folds, $dN = - H dt$:\footnote{The parameters $s$ and $\tilde s$ correspond to
the parameters $\kappa$ and $\tilde\kappa$ in Bean, {\it et al.} \cite{Bean:2008ga}}.
\begin{eqnarray}
\label{epsN}
\epsilon &\equiv& \frac{1}{H} \frac{d H}{dN}
,\\ \label{sN}
s &\equiv& -\frac{1}{c_S} \frac{d c_S}{dN},\\ \label{stilN}
\tilde{s} &\equiv& \frac{1}{{\cal{L}}_{X}}\frac{d {\cal{L}}_{X}}{dN}.
\end{eqnarray}
The Friedmann equation can be written in terms of the reduced Planck mass  $M_P=1/\sqrt{8\pi G}$
\begin{equation}
H^2 = \frac{1}{3 M_P^2} \rho = \frac{1}{3 M_P^2} \left(2X {\cal
L}_X - {\cal L}\right),
\end{equation}
and the continuity equation is
\begin{equation}
\dot\rho = 2 H {\dot H} = -3 H \left(\rho + p\right) = - 6 H X {\cal L}_X.
\end{equation}
For monotonic field evolution, the field value $\phi$ can be used as a ``clock'', and all other quantities expressed as functions of $\phi$, for
example $X = X\left(\phi\right)$, ${\cal L} = {\cal L}\left[X\left(\phi\right),\phi\right]$, and so on. We consider the homogeneous case, so that
$\dot\phi = \sqrt{2 X}$. Using
\begin{equation}
\frac{d}{dt} = \dot\phi \frac{d}{d \phi} = \sqrt{2 X} \frac{d}{d\phi},
\end{equation}
we can re-write the Friedmann and continuity equations as the {\it Hamilton Jacobi} equations,
\begin{eqnarray}
\label{hamjac1} \dot
\phi = \sqrt{2 X} &=& -\frac{2M_P^2}{{\cal L}_{X}}H'(\phi),\\ \label{hamjac2}
3M_P^2H^2(\phi)&=&\frac{4M_P^4{H'\left(\phi\right)}^2}{{\cal L}_{X}}-{\cal L}.
\end{eqnarray}
where a prime denotes a derivative with respect to the field $\phi$. The number of e-folds $dN$ can similarly be re-written in terms of $d\phi$ by:
\begin{eqnarray}
\label{eq:Nphi}
dN \equiv -H dt &&= - \frac{H}{\sqrt{2 X}} d\phi\\
&&= \frac{{\cal L}_X}{2 M_P^2} \left(\frac{H\left(\phi\right)}{H'\left(\phi\right)}\right) d\phi.
\end{eqnarray}
The flow parameters $\epsilon$, $s$, and $\tilde s$ (\ref{epsN}) can therefore be written as derivatives with respect to the field $\phi$ as:
\begin{eqnarray}
\label{defeps}
\epsilon\left(\phi\right) &=& \frac{2
M_P^2}{{\cal{L}}_{X}}
\left(\frac{H'\left(\phi\right)}{H\left(\phi\right)}\right)^2,\\ \label{defs}
s\left(\phi\right) &=& - \frac{2
M_P^2}{{\cal{L}}_{X}}
\frac{H'\left(\phi\right)}{H\left(\phi\right)} \frac{c_S'\left(\phi\right)}{c_S\left(\phi\right)},\\ \label{defstil}
\tilde{s}\left(\phi\right) &=& \frac{2
M_P^2}{{\cal{L}}_{X}}
\frac{H'\left(\phi\right)}{H\left(\phi\right)}\frac{{\cal{L'}}_{X}}{{\cal{L}}_{X}}.
\end{eqnarray}
Taking successive derivatives $d/dN$ with respect to the number of e-folds yields an infinite hierarchy of flow equations
\cite{Kinney:2002qn,Bean:2008ga},
\begin{eqnarray}
\label{flowequations} \frac{d \epsilon}{d N} &=&
-\epsilon\left(2 \epsilon - 2 \tilde{\eta} + \tilde{s}\right),\cr
\frac{d \tilde{\eta}}{d
N} &=& -\tilde{\eta}\left(\epsilon + \tilde{s}\right) + {}^2 \lambda,\cr
\frac{d s}{d N} &=& -s \left(\epsilon -
\tilde{\eta}+\tilde{s}+s\right) + \epsilon \rho,\cr
\cr\frac{d \tilde{s}}{d N} &=& -\tilde{s} \left(\epsilon -
\tilde{\eta}+2 \tilde{s} \right) + \epsilon {}^1 \beta,
\cr \frac{d
{}^\ell \lambda}{d N} &=& -{}^\ell \lambda \left[\ell \left(\tilde{s} +
\epsilon\right) - \left(\ell - 1\right) \tilde{\eta}\right] + {}^{\ell +
1}\lambda,
\cr \frac{d {}^\ell \alpha}{d N} &=& -{}^\ell \alpha
\left[\left(\ell - 1\right)
(\epsilon-\tilde{\eta})+\ell\tilde{s} + s\right] + {}^{\ell + 1} \alpha,\cr
\frac{d {}^\ell \beta}{d N} &=& -{}^\ell \beta
\left[\left(\ell - 1\right)
(\epsilon-\tilde{\eta})+\left(\ell+1\right)\tilde{s}\right] + {}^{\ell + 1} \beta,
\end{eqnarray}
where the higher-order flow parameters are defined as follows, where $\ell = 1,\ldots\infty$ is an integer parameter:
\begin{eqnarray}
\label{hflowpar}
\tilde{\eta}\left(\phi\right) &=& {}^1 \lambda = \frac{2
M_P^2}{{\cal{L}}_{X}}
\frac{H''\left(\phi\right)}{H\left(\phi\right)},\cr
{}^\ell \lambda\left(\phi\right) &=& \left(\frac{2
M_P^2}{{\cal{L}}_{X}}\right)^{\ell}
\left(\frac{H'\left(\phi\right)}{H\left(\phi\right)}\right)^{\ell
- 1} \frac{1}{H\left(\phi\right)} \frac{d^{\ell + 1}
}{d \phi^{\ell + 1}}H\left(\phi\right),\cr
{}^\ell
\alpha\left(\phi\right) &=& \left(\frac{2
M_P^2}{{\cal{L}}_{X}}\right)^{\ell}
\left(\frac{H'\left(\phi\right)}{H\left(\phi\right)}\right)^{\ell
- 1} \frac{1}{c_S^{-1}(\phi)} \frac{d^{\ell + 1}
}{d \phi^{\ell + 1}}c_S^{-1}(\phi),
\cr {}^\ell
\beta\left(\phi\right) &=& \left(\frac{2
M_P^2}{{\cal{L}}_{X}}\right)^{\ell}
\left(\frac{H'\left(\phi\right)}{H\left(\phi\right)}\right)^{\ell
- 1} \frac{1}{{\cal{L}}_{X}} \frac{d^{\ell + 1}
}{d \phi^{\ell + 1}}{\cal{L}}_{X}.
\end{eqnarray}
Solutions to this infinite hierarchy of flow equations are equivalent to solutions of the scalar field equation of motion. In the next section, we
specialize to the case where the  flow parameters are constant, which results in an exactly solvable system.

\section{Cosmological solutions for constant flow parameters}
\label{sec:flowsolutions}

The simplest way to solve the flow equations derived in the preceding section is to take all of the flow parameters to be constant,
\begin{equation}
\frac{d \epsilon}{d N} = \frac{d s}{d N} = \frac{d \tilde s}{d N} = \frac{d {}^\ell\lambda}{dN} = \frac{d {}^\ell\alpha}{dN}  = \frac{d {}^\ell\beta}{dN} = 0.
\end{equation}
Then, from (\ref{epsN}-\ref{stilN}) we easily find the following relations:
\begin{eqnarray}
\label{solflowparN}
H &\propto& e^{\epsilon N}
,\cr
c_S  &\propto&  e^{-s N},\cr
{\cal{L}}_{X} &\propto& e^{\tilde{s}N},
\end{eqnarray}
The first two are identical to the DBI case, Eqs.
(\ref{eq:DBIsoundspeed}) and (\ref{eq:Hphi}), but in the fully
general case ${\cal L}_X$ evolves independently of $c_S$. It is
straightforward to verify that the full flow hierarchy
(\ref{hflowpar}) reduces to an exactly solvable set of algebraic
equations, with the higher-order parameters expressed in terms of
$\epsilon$, $s$, and $\tilde s$. We can use the relations
(\ref{defeps},\ref{defs},\ref{defstil}) to solve for
$H\left(\phi\right)$, $c_S\left(\phi\right)$, and ${\cal
L}_X\left(\phi\right)$ as follows: from Eqs.
(\ref{defeps},\ref{defstil}), we have
\begin{equation}
{\tilde s} = \frac{2 M_P^2}{{\cal L}_X} \left(\frac{H'}{H}\right)\frac{{\cal L}_X'}{{\cal L}_X} = M_p \sqrt{2 \epsilon} \frac{{\cal L}_X'}{{\cal L}_X} = {\rm const.}
\end{equation}
We then have a differential equation for ${\cal L}_X$,
\begin{equation}
\frac{{\cal{L'}}_{X}}{{\cal{L}}_{X}^{3/2}}=\frac{\tilde{s}}{
M_P \sqrt{2 \epsilon}} = {\rm const.},
\end{equation}
with solution
\begin{equation}
\label{lagXphi}
{\cal{L}}_{X}\left(\phi\right)=\frac{8\epsilon}{\tilde{s}^2}\left(\frac{M_P}{\phi}\right)^{2},
\end{equation}
where the integration constant has been absorbed into a field
redefinition. From Eq. (\ref{solflowparN}), the field then evolves
as
\begin{equation}
\phi^2 \propto e^{-{\tilde s} N},
\end{equation}
so that the direction of the field evolution depends on the sign of $\tilde s$,
\begin{equation}
\frac{d \phi}{\phi} = -\frac{\tilde s}{2} d N.
\end{equation}
Equation (\ref{defeps}) then reduces to
\begin{equation}
\left(\frac{H'}{H}\right)^2 = \frac{4 \epsilon^2}{{\tilde s}^2 \phi^2},
\end{equation}
with solution
\begin{equation}
H \propto \phi^{\pm 2 \epsilon / \tilde s}.
\end{equation}
The sign ambiguity can be resolved by requiring that the universe be expanding, $d H / d N > 0$, so that
\begin{equation}
H \propto \phi^{-2 \epsilon / {\tilde s}} \propto e^{\epsilon N}.
\end{equation}
Finally, we solve for the speed of sound using Eq. (\ref{defs}), which reduces to
\begin{equation}
\frac{c_S'}{c_S} = \frac{2 s}{\tilde s} = {\rm const.},
\end{equation}
with solution
\begin{equation}
c_S \propto \phi^{2 s / \tilde s}.
\end{equation}
Since our choice of $N = 0$ corresponds to an arbitrary renormalization of the scale factor $a \propto e^{-N}$, we can without loss of generality
define $c_S = 1$ at $N = 0$, so that the general solution for the background evolution is given by
\begin{equation}
\label{sqrtlagX}
{\cal{L}}_{X} = \frac{8\epsilon}{{\tilde s}^2}\left(\frac{M_P}{\phi}\right)^2,
\end{equation}
\begin{equation}
\label{hubpar}
H\left(\phi\right)=H_0\left(\frac{\phi}{\phi_0}\right)^{-2\epsilon/\tilde{s}},
\end{equation}
\begin{equation}
\label{gamm}
c_S\left(\phi\right)=\left(\frac{\phi}{\phi_0}\right)^{2s/\tilde{s}},
\end{equation}
where the field evolves as
\begin{equation}
\label{efoldsphi}
\frac{\phi}{\phi_0} = e^{-{\tilde s} N / 2}.
\end{equation}

We can derive the time dependence of the scale factor using the Hamilton-Jacobi equation (\ref{hamjac1}),
\begin{eqnarray}
\label{dotphi} \dot
\phi&=&\frac{\tilde{s}}{2} H\left(\phi\right) \phi = \sqrt{2 X},
\end{eqnarray}
so that the kinetic term can be written as
\begin{eqnarray}
\label{Xphi}
X\left(\phi\right)&=\frac{\tilde{s}^2}{8} H^2\left(\phi\right) \phi^2 .
\end{eqnarray}
Integrating expression (\ref{dotphi}) gives
\begin{equation}
\label{hubpart}
H\left(t\right)=\frac{1}{\epsilon t},
\end{equation}
so that the scale factor evolves as a power-law in time, consistent with the relation (\ref{eq:defofepsilon}) between $\epsilon$ and the equation of
state $w = p / \rho$,
\begin{equation}
\label{scalefac}
a\left(t\right)\propto t^{1/\epsilon} = t^{2 / 3 \left(1 + w\right)}.
\end{equation}
Radiation-dominated evolution therefore corresponds to $\epsilon = 2$, and matter-dominated evolution corresponds to $\epsilon = 3/2$. Inflation
corresponds to $\epsilon < 1$. The comoving Hubble horizon evolves proportional the to the conformal time,
\begin{equation}
d_H \propto (a H)^{-1} \propto e^{\left(1 - \epsilon\right) N} \propto \tau,
\end{equation}
and the acoustic horizon evolves as
\begin{equation}
D_H \propto \frac{c_S}{a H} \propto e^{\left(1 - \epsilon - s\right) N} \propto \tau^{\left(1 - \epsilon - s\right) / \left(1 - \epsilon\right)},
\end{equation}
identically to the DBI case discussed in Sec. \ref{sec:TachyacousticCosmology}. For $\epsilon > 1$, the acoustic horizon is shrinking in comoving units
for $s < 1 - \epsilon$. Note that this behavior is independent of the parameter $\tilde s$, which determines the form of the Lagrangian, as we discuss
in the next section.

\section{Reconstructing the action}
\label{sec:reconstruction}

In the past two sections we have solved the flow hierarchy for a
model characterized by constant flow parameters, which allowed us
to solve for $H\left(\phi\right)$, $c_S\left(\phi\right)$, and
${\cal L}_X\left(\phi\right)$; only the derivative of the
Lagrangian with respect to the kinetic term $X$ is determined.
Therefore this solution corresponds not to a single action but a
class of actions. In this section we derive a general equation for
Lagrangians in this class, and discuss two specific examples.

From Eqs. (\ref{lagXphi}) and (\ref{gamm}), we see that the speed of sound $c_S$ can be written in terms of ${\cal{L}}_{X}$
\begin{equation}
\label{lagXgamma}
c_S^2=C^{-1} {\cal{L}}_{X}^{-2 s/\tilde{s}} = \left[1 + 2 X \frac{{\cal L}_{XX}}{{\cal L}_X}\right]^{-1},
\end{equation}
where we have used Eq. (\ref{defspeedofsound}), and defined
\begin{equation}
\label{defK}
C \equiv \left(\frac{\tilde{s}^2\phi_0^2}{8M_P^2\epsilon}\right)^{2 s / \tilde s}.
\end{equation}
The result is a differential equation for the function
${\cal{L}}\left(X,\phi\right)$:
\begin{equation}
\label{difflag}
2X{\cal{L}}_{XX}+{\cal{L}}_X-C {\cal{L}}_{X}^{n}=0,
\end{equation}
where we have defined
\begin{equation}
\label{difflagconst}
n\equiv 1 + \frac{2 s}{\tilde{s}}.
\end{equation}
Therefore, by specifying a relationship between the parameters $s$ and $\tilde s$, we can construct a Lagrangian as the solution to the differential
equation (\ref{difflag}). For example, a canonical Lagrangian with speed of sound $c_S = {\rm const.} = 1$ is just the case $s = 0$, so that $n = 1$
and $C = 1$, and Eq. (\ref{difflag}) becomes
\begin{equation}
{\cal{L}}_{XX}=0,
\end{equation}
with general solution
\begin{equation}
{\cal L} = f\left(\phi\right) X - V\left(\phi\right).
\end{equation}
Here $f\left(\phi\right)$ and $V\left(\phi\right)$ are free functions which arise from integration of the second-order equation (\ref{difflag}).
The function $f\left(\phi\right)$ can be eliminated by a field redefinition $d \varphi = \sqrt{f\left(\phi\right)} d \phi$, resulting in a manifestly
canonical Lagrangian for $\varphi$, as we would expect from setting $c_S = 1$. A canonical Lagrangian can support inflationary solutions, but not
tachyacoustic solutions, and is therefore not of interest here. However, other choices of $n$ do yield tachyacoustic solutions, and we focus on two
such choices:
\begin{enumerate}
\item{$n=0$: A Cuscuton-like model.}
\item{$n=3$: A DBI model.}
\end{enumerate}
We discuss each case separately below.

\subsection{{$\mathbf{n=0}$\bf{: A Cuscuton-like model}}}
\label{subsec:cuscutonlike}

The case $n=0$ corresponds to $\tilde{s}=-2s$ in (\ref{difflagconst}), with solution
\begin{equation}
\label{lagcuscs}
{\cal{L}}\left(X,\phi\right)=2f\left(\phi\right)\sqrt{X} + C X - V\left(\phi\right).
\end{equation}
This Lagrangian is similar to a ``cuscuton'' Lagrangian \cite{Afshordi:2006ad}, with the addition of a term proportional to $X$.  Unlike the original
Cuscuton model, which represents a causal field
with infinite speed of sound, the solution obtained here is valid for the general case, in which the speed of sound can be finite. A similar cuscuton-like Lagrangian was considered in Ref. \cite{Piao:2008ip}.

As in the canonical case, the functions $f\left(\phi\right)$ and $V\left(\phi\right)$ are free functions resulting from integrating Eq. (\ref{difflag}). Unlike the canonical case, however, neither can be removed by a field redefinition. However, both functions are fully determined by our choice of solution with $\epsilon$, $s$, and $\tilde s$ constant. Differentiating Eq. (\ref{lagcuscs}) with respect to $X$ gives
\begin{equation}
{\cal L}_X = \frac{f\left(\phi\right)}{\sqrt{X}} + C = \frac{2 \epsilon}{s^2} \left(\frac{M_P}{\phi}\right)^2,
\end{equation}
where the right hand side is the solution (\ref{sqrtlagX}). Then
\begin{eqnarray}
f\left(\phi\right) &=& \sqrt{X} \left(\frac{2 \epsilon}{s^2}\right) \left(\frac{M_P}{\phi_0}\right)^2 \left[\left(\frac{\phi_0}{\phi}\right)^2 - 1\right]\cr
&=& \sqrt{X} \left(\frac{2 \epsilon}{s^2}\right) \left(\frac{M_P}{\phi_0}\right)^2 \left[c_S^2\left(\phi\right) - 1\right]
\end{eqnarray}
where for $2\tilde{s} = - s$, the expression (\ref{gamm}) for the
speed of sound becomes
\begin{equation}
\label{cscusc}
c_S\left(\phi\right)= \left(\frac{\phi_0}{\phi}\right).
\end{equation}
The Lagrangian (\ref{lagcuscs}) can then be written as
\begin{equation}
{\cal L} = X \left(\frac{2 \epsilon}{s^2}\right) \left(\frac{M_P}{\phi_0}\right)^2 \left[2 c_S^2\left(\phi\right) - 1\right] - V\left(\phi\right).
\end{equation}
The Hubble parameter (\ref{hubpar}) is given by
\begin{equation}
\label{eq:Hcusc}
H\left(\phi\right) = H_0 \left(\frac{\phi}{\phi_0}\right)^{\epsilon / s},
\end{equation}
and we can then express the kinetic term as a function of $\phi$ using Eq. (\ref{Xphi}):
\begin{equation}
X\left(\phi\right) = \frac{s^2}{2} H^2 \phi^2 = \frac{s^2}{2} \frac{ \phi_0^2 H^2\left(\phi\right)}{c_s^2\left(\phi\right)},
\end{equation}
The Lagrangian (\ref{lagcuscs}) can then be written entirely as a function of the field $\phi$,
\begin{equation}
{\cal L} = M_P^2 \epsilon H^2\left(\phi\right) \left[2 - \frac{1}{c_S^2\left(\phi\right)}\right] - V\left(\phi\right).
\end{equation}
The Hamilton-Jacobi Equation (\ref{hamjac1}) becomes:
\begin{eqnarray}
3 M_P^2 H^2 &=& 2 M_P^2 \epsilon H^2 - {\cal L}\cr
&=& V\left(\phi\right) + \frac{M_P^2 \epsilon H^2}{c_S^2},
\end{eqnarray}
and we have an expression for the potential $V\left(\phi\right)$,
\begin{equation}
V\left(\phi\right) = M_P^2 H^2\left(\phi\right) \left[3 - \frac{\epsilon}{c_S^2\left(\phi\right)}\right].
\end{equation}
The Hubble parameter $H\left(\phi\right)$ and the speed of sound $c_S\left(\phi\right)$ are given by Eqs. (\ref{eq:Hcusc}) and (\ref{cscusc}),
respectively. For $\phi / \phi_0 \ll 1$, the speed of sound is much greater than the speed of light, $c_S \gg 1$, and the potential is approximately
\begin{equation}
V\left(\phi\right) \simeq 3 M_P^2 H^2\left(\phi\right) = 3 M_P^2 H_0^2 \left(\frac{\phi}{\phi_0}\right)^{2 \epsilon / s},
\end{equation}
which can be recognized as a slow-roll-like solution dominated by
the potential $H^2 \simeq V^2 / {3 M_P^2}$. For $s < 0$, the
field is rolling away from the origin, and for $s < 1 - \epsilon$
the comoving acoustic horizon is shrinking and the solution is
tachyacoustic.

\subsection{{$\mathbf{n=3}$\bf{: The DBI model}}}
\label{subsec:dbi}

The case $n=3$, corresponds to $\tilde{s}=s$; then, from (\ref{defs}) and (\ref{defstil}), we find ${\cal{L}}_{X}=c_S^{-1}$. Eq. (\ref{difflagconst})
is then
\begin{equation}
c_S^2=\frac{1}{C {\cal{L}}_{X}^2},
\end{equation}
so that we can take $C = 1$ without loss of generality. Therefore, the Lagrangian assumes the well-known DBI form,
\begin{equation}
\label{lagDBIs}
{\cal{L}}\left(X,\phi\right)=-f^{-1}\left(\phi\right)\sqrt{1-f\left(\phi\right)X} + f^{-1}\left(\phi\right) - V\left(\phi\right).
\end{equation}
The DBI model with constant flow parameters is extensively discussed in Ref. \cite{Kinney:2007ag}, and the reader is referred this paper for further
details. For $\epsilon$ and $s$ constant, the functions $V$ and $f$ are fully determined and are given by
\begin{eqnarray}
\label{eq:potentials}
V\left(\phi\right) &=& 3 M_P^2 H^2\left(\phi\right) \left[1 - \left(\frac{2 \epsilon}{3}\right) \frac{1}{1 + c_S\left(\phi\right)}\right],\cr
f\left(\phi\right) &=& \left(\frac{1}{2 M_P^2 \epsilon}\right) \frac{1 - c_S^2\left(\phi\right)}{H^2\left(\phi\right) c_S\left(\phi\right)}.
\end{eqnarray}
The Hubble parameter and speed of sound are given by:
\begin{equation}
H\left(\phi\right) = H_0 \left(\frac{\phi}{\phi_0}\right)^{- 2 \epsilon / s},
\end{equation}
and
\begin{equation}
c_S\left(\phi\right) = \left(\frac{\phi}{\phi_0}\right)^2.
\end{equation}
DBI Lagrangians allow for either inflationary or tachyacoustic evolution \cite{Magueijo:2008sx}, depending on the values of $\epsilon$ and $s$. Note that for $c_S > 1$, the
function $f$ is negative, which has consequences for embedding such a model in string theory, which we discuss in Sec. \ref{sec:Conclusions}.

In this section, we have explicitly constructed Lagrangians,
including fully determined potentials, for which the flow
parameters are constant and the background evolution can be solved
exactly. For suitable choices of the flow parameters, the
evolution is tachyacoustic, {\it i.e.} with a growing comoving
Hubble horizon and a shrinking comoving acoustic horizon. In the
next section, we discuss the generation of curvature perturbations
at the acoustic horizon and show that such perturbations are
nearly scale-invariant, consistent with observation.

\section{Cosmological Perturbations for constant flow parameters}
\label{sec:Perturbations}

We can deal with cosmological perturbations in this general k-essence model with constant flow parameters in the same way as performed in
\cite{Kinney:2007ag}. Following the approach of Garriga and Mukhanov \cite{Garriga:1999vw} we start with the
perturbed Einstein equations,
\begin{eqnarray}
\label{eq:perteinst}
\frac{d}{dt}\left(\frac{\delta\phi}{\dot{\phi}}\right)&=&\Phi+\frac{2M_P^2c_S^2}{a^2(\rho+p)}\nabla^2\Phi\nonumber \\
\frac{d}{dt}\left(a\Phi\right)&=&\frac{a(\rho+p)}{2M_P^2}\left(\frac{\delta\phi}{\dot{\phi}}\right),
\end{eqnarray}
where $\Phi$ is the Bardeen potential and $\delta\phi$ is the
perturbation of the field $\phi$. Equations
(\ref{eq:perteinst}) can be cast into a more convenient form by
changing the perturbations $\Phi$ and $\delta\phi$ to the new
variables $\zeta$ and $\xi$ defined by
\begin{eqnarray}
\label{eq:newpert}
\xi&=&\frac{2M_P^2\Phi a}{H}\nonumber \\
\zeta&=&H\frac{\delta\phi}{\dot{\phi}}+\Phi,
\end{eqnarray}
so that the perturbed Einstein equations (\ref{eq:perteinst}) become
\begin{eqnarray}
\label{eq:newperteinst}
\dot\xi&=&\frac{a(\rho+p)}{H^2}\zeta,\nonumber \\
\dot\zeta&=&\frac{c_S^2H^2}{a^3(\rho+p)}\nabla^2\xi.
\end{eqnarray}
As usual, we introduce a new variable $z$ and the gauge-invariant Mukhanov potential $u$ as
\begin{equation}
\label{eq:defz}
z=\frac{a(\rho+p)^{1/2}}{c_SH},~~~~~ u=z\zeta;
\end{equation}
then, from (\ref{eq:newperteinst}) we derive the mode equation for $u(\tau)\propto u_k(\tau)\exp(i\mathbf{k}\cdot\mathbf{x})$, given by
\begin{equation}
\label{eq:modeeq}
u_k'' - \left[\left(c_S k\right)^2 + \frac{z''}{z}\right] u_k = 0,
\end{equation}
where a prime denotes a derivative with respect to conformal time,
$ds^2 = a^2\left(\tau\right) \left(d \tau^2 - d {\bf x}^2\right)$.
It is easy to show that the variable $z$, defined by
(\ref{eq:defz}) can be cast into the following form,
\begin{equation}
\label{newdefz}
z=-\frac{aM_P\sqrt{2\epsilon}}{c_S};
\end{equation}
then, using
\begin{equation}
\label{ddtau}
\frac{d}{d\tau}=-aH\frac{d}{dN},
\end{equation}
we can evaluate the ratio $z''/z$ in (\ref{eq:modeeq}) in terms of the flow parameters (\ref{defeps}-\ref{hflowpar}); the result is
\begin{equation}
\label{zF}
\frac{z''}{z} = a^2 H^2
\bar F\left(\epsilon,\tilde{\eta},s,\tilde{s},{}^2\lambda,{}^1\alpha,{}^1\beta\right),
\end{equation}
where
\begin{eqnarray}
\label{defF}\bar F &\equiv & 2+2\epsilon-3\tilde{\eta}-3s+\frac{3}{2}\tilde{s}+ 2\epsilon^2+\frac{5}{4}\tilde{s}^2-2s\tilde{s}\nonumber \\
&+&\tilde{\eta}^2+2\epsilon(\tilde{s}-s)+3\tilde{\eta} s-\frac{5}{2}\tilde{\eta}\tilde{s}
-4\tilde{\eta}\epsilon+ {}^2 \lambda\nonumber \\ &-&\frac{1}{2}\epsilon \left({}^1 \alpha\right)+\epsilon \left({}^1 \beta\right).
\end{eqnarray}
Next, it is convenient to change the conformal time, $\tau$, to the ratio of wavenumber to the sound horizon,
\begin{equation}
y \equiv \frac{c_Sk}{a H};
\end{equation}
then, conformal time derivatives switch to
\begin{equation}\label{dery1}
\frac{d}{d\tau} = - aH \left(1 - \epsilon -
s\right)y \frac{d}{dy},
\end{equation}
and
\begin{equation}\label{dery}
\frac{d^2}{d\tau^2} = a^2 H^2 \left[\left(1 - \epsilon -
s\right)^2 y^2 \frac{d^2}{dy^2} +
\bar G\left(\epsilon,\tilde{\eta},s,\tilde{s},{}^1 \alpha\right) y \frac{d}{d y}\right],
\end{equation}
where
\begin{eqnarray}
\label{defG}\bar G &\equiv& -s+\epsilon(2s+\tilde{s})+s(2s+\tilde{s})+2\epsilon^2-2\epsilon\tilde{\eta}-s\tilde{\eta}\nonumber \\ &-&\epsilon \left({}^1 \alpha\right).
\end{eqnarray}
It is important to stress that the functions $F$ and $G$ derived above hold in general; they reduce to the well known expressions in the DBI limit
\cite{Kinney:2007ag}, which, in this case, $s=\tilde{s}$ and ${}^1 \alpha={}^1 \beta=\rho$. Substituting (\ref{zF}) and (\ref{dery}) into the mode
equation (\ref{eq:modeeq}), we find
\begin{equation}
\label{exactmode} \left(1 - \epsilon -s\right)^2 y^2 \frac{d^2
u_k}{dy^2} + \bar G y \frac{d u_k}{d y} + \left[y^2 - \bar F\right] u_k = 0,
\end{equation}
which is an exact equation, without any assumption of slow-roll.

In the case where the flow parameters are constant, we can use the differential equations (\ref{flowequations}) to reduce the number of independent
parameters. We have
\begin{eqnarray}
\tilde{\eta} &=& \frac{1}{2}\left(2 \epsilon + \tilde{s}\right),\cr {}^2 \lambda
&=& \frac{1}{2}\left(2 \epsilon + \tilde{s}\right) \left(\epsilon +
\tilde{s}\right),\cr {}^1 \alpha &=&
\frac{s}{2 \epsilon}\left(2s +\tilde{s}\right),\cr
{}^1 \beta &=& \frac{3 {\tilde{s}}^2}{2 \epsilon};
\end{eqnarray}
then, substituting these values into expressions (\ref{defF}) and (\ref{defG}), we find, respectively,
\begin{eqnarray}
\label{Fflowc}\bar F &=& 2-\epsilon-3s+\frac{9}{4}\tilde{s}^2-\frac{3}{4}s\tilde{s}+\epsilon s-\frac{1}{2}s^2,
\end{eqnarray}
\begin{eqnarray}
\label{Gflowc}\bar G &=& s(-1+\epsilon+s).
\end{eqnarray}
It is important to notice that $\bar F$ is different from the
corresponding expression found in the DBI case
\cite{Kinney:2007ag}, since the gauge-dependent $\tilde{s}$ comes
into play. However, $\bar G$ is identical to its DBI analog, and
it is expected since basically it comes from the change of
variables $\tau\rightarrow y$, which depends solely on the
parameters $c_S$ and $H$, and not on ${\cal{L}}_{X}$. For constant
flow parameters we can solve equation (\ref{exactmode}) exactly,
and the solutions are given by
\begin{eqnarray}
\label{uksol}
u_{k}(y)&=&y^{\frac{1-\epsilon}{2(1-\epsilon-s)}}\left[c_{1}H_{\nu}^{(1)}\left(\frac{y}{1-\epsilon-s}\right)\right.\nonumber \\
&+&\left.c_{2}H_{\nu}^{(2)}\left(\frac{y}{1-\epsilon-s}\right)\right],
\end{eqnarray}
where $c_1$ and $c_2$ are constants, and $H_{\nu}^{(1)}$, $H_{\nu}^{(2)}$ are Hankel functions of first and second kind, respectively. The order $\nu$
of the Hankel function is given by
\begin{eqnarray}
\label{horder}
\nu^2=\frac{9-6\epsilon-12s+9\tilde{s}^2-3s\tilde{s}+4\epsilon s-2s^2+\epsilon^2}{4(1-\epsilon-s)^2};
\end{eqnarray}
next, using (\ref{solflowparN}), (\ref{ddtau})
and (\ref{dery1}) we find that
\begin{eqnarray}
\label{csy}
c_S\propto y^{s/(\epsilon+s-1)};
\end{eqnarray}
then, imposing the Bunch-Davies vacuum $c_2=0$, and normalizing the mode amplitudes by means of the canonical quantization condition
\begin{equation}
\label{Wronskian}
u_{k}^{*}\frac{du_{k}}{dy}-u_{k}\frac{du_{k}^{*}}{dy}=\frac{i}{c_{s}k(1-\epsilon-s)},
\end{equation}
we find
\begin{equation}
\label{uksolution}
u_{k}(y)=\frac{1}{2}\sqrt{\frac{\pi}{c_{s}k}}\sqrt{\frac{y}{1-\epsilon-s}}H_{\nu}\left(\frac{y}{1-\epsilon-s}\right),
\end{equation}
which differs from the DBI case only in the order of the Hankel function (\ref{horder}). In the small wavelength limit $y\rightarrow\infty$
the early-time behavior of $u_k$ will be identical to DBI \cite{Kinney:2007ag} for constant flow parameters
\begin{equation}
\label{uksmall}
u_{k}=\frac{1}{\sqrt{2c_{s}k}}e^{iy/(1-\epsilon-s)},
\end{equation}
whereas in the late-time behavior $y\rightarrow0$ the mode function behaves as
\begin{equation}
\label{uklong}
\left|u_{k}(y)\right| \rightarrow
2^{\nu-3/2}\frac{\Gamma(\nu)}{\Gamma(3/2)}(1-\epsilon-s)^{\nu-1/2}
\frac{y^{1/2-\nu}}{\sqrt{2 c_{s}k}}.
\end{equation}
From (\ref{uklong}) we can derive the expression for the scalar spectral index $n_s$. Using the definition of the power spectrum of curvature
perturbations
\begin{equation}
\label{curvspec1}
P_{\cal R}\left(k\right) =
\frac{k^3}{2 \pi^2} \left\vert\frac{u_k}{z}\right\vert^2,
\end{equation}
and substituting expressions (\ref{newdefz}) and (\ref{uklong}) into (\ref{curvspec1}), we find
\begin{equation}
\label{curvspec2}
P_{\mathcal{R}}=\frac{\left\vert f(\nu)\right\vert^2}{8\pi^2M_P^2}\frac{H^2}{c_S\epsilon}
\end{equation}
at horizon crossing, where $f(\nu)$ is a constant given by
\begin{equation}
\label{fnu}
f(\nu)=2^{\nu-3/2}\frac{\Gamma(\nu)}{\Gamma(3/2)}(1-\epsilon-s)^{\nu-1/2};
\end{equation}
then, from the definition of the scalar spectral index
\begin{equation}
\label{defns} n_{s}-1 \equiv
\frac{d\ln P_{\mathcal{R}}}{d\ln k},
\end{equation}
and using
\begin{equation}
\label{ddk} \frac{d}{d\ln
k}=-\left(\frac{1}{1-\epsilon-s}\right)\frac{d}{dN},
\end{equation}
we see that the spectral index $n_s$ assumes the form
\begin{equation}
\label{specindex} n_{s}=1-\frac{2\epsilon+s}{1-\epsilon-s},
\end{equation}
which does not depend on the gauge-dependent parameter $\tilde{s}$, and is identical to its DBI analog. This is expected since the power spectrum
evaluated at the horizon crossing, equation (\ref{curvspec2}), depends solely on $H$ and $c_S$, whose derivatives with respect to $N$ are related
to the gauge-invariant flow parameters $\epsilon$ and $s$. The scale-invariant limit is $s = - 2 \epsilon$.

\section{Conclusions}
\label{sec:Conclusions}

In this paper we have demonstrated that accelerated expansion or a
collapsing universe are not the only ways to dynamically generate
a scale-invariant spectrum of superhorizon curvature
perturbations. There is a third way: a superluminal acoustic cone
which is shrinking in comoving coordinates. Curvature
perturbations generated at the acoustic horizon are familiar from
inflationary scenarios based on non-canonical Lagrangians such as
$k$-inflation and DBI inflation. Such non-canonical Lagrangians
arise naturally in string theory.  However, in these scenarios,
{\it both} the Hubble horizon and the acoustic horizon are
shrinking in comoving units, and the acoustic horizon is typically
smaller than the Hubble horizon, {\it i.e.} $c_S < 1$. It is
natural to ask whether tachyacoustic models have a similar,
natural stringy embedding, especially since the DBI action
(\ref{eq:DBIlagrangian}) naturally admits tachyacoustic solutions.
Such an embedding is nontrivial, however, since the frequently
considered case of a 3+1 dimension d-brane evolving in a
higher-dimensional throat is ill-defined in the $c_S > 1$ limit.
To see this, consider the full ten-dimensional metric of throat
plus brane \cite{Klebanov:2000hb},
\begin{equation}
\label{eq:DBImetric}
ds^2_{10} = h^2\left(r\right) ds^2_4 + h^{-2}\left(r\right) \left(d r^2 + r^2 ds^2_{X_5}\right).
\end{equation}
The field $\phi$ is simply related to the coordinate in the throat
$r$ as $\phi = \sqrt{T_3} r$, where the brane tension  $T_3$
depends on the string scale $m_s$ and the string coupling $g_s$ as
\cite{Lidsey:2007gq}
\begin{equation}
T_3 = \frac{m_s^4}{\left(2 \pi\right)^3 g_s}.
\end{equation}
The Lagrangian for the field $\phi$ can be shown to be of the DBI form (\ref{eq:DBIlagrangian}),
where the inverse brane tension $f\left(\phi\right)$ is given in terms of the warp factor $h\left(\phi\right)$ by
\begin{equation}
\label{eq:warpfactor}
f\left(\phi\right) = \frac{1}{T_3 h^4\left(\phi\right)}.
\end{equation}
The problem is immediately evident: superluminal propagation $c_S
> 1$ requires $f < 0$, so that the factor $h^2\left(\phi\right)$
appearing in the metric (\ref{eq:DBImetric}) is imaginary, and the
metric is ill-defined. Therefore, although the DBI action itself
admits tachyacoustic solutions, this limit does not correspond to
a well-defined string solution. It is not clear whether or not
string manifolds exist which self-consistently admit solutions
with $c_S > 1$.

We calculate the scalar spectral index of perturbations for tachyacoustic solutions, and find
\begin{equation}
n_{s}=1-\frac{2\epsilon+s}{1-\epsilon-s}.
\end{equation}
Unlike inflationary models, radiation-dominated tachyacoustic
models do not require a period of explosive entropy production to
transition to a ``hot'' Big Bang cosmology. The early universe
must be scalar-field dominated, but the temperature of the
universe is not driven exponentially to zero, since the scalar has
a radiation equation of state at all times, and entropy density is
conserved (for any radiation component with density $\rho_\gamma$,
the ratio $\rho_\phi / \rho_{\gamma} = {\rm const}$). The scalar field $\phi$ must eventually decay to Standard Model degrees of freedom, but as long as this happens before primordial nucleosynthesis, the model will match observations. A slow or late decay of $\phi$ into other degrees of freedom would also suppress the production of unwanted relics such as monopoles or gravitinos. For
radiation-dominated tachyacoustic expansion with $\epsilon = 2$,
the spectral index is
\begin{equation}
n = 1 + \frac{4 + s}{1 + s},
\end{equation}
where we have $s < -3$ for a shrinking comoving acoustic cone. For
$s < -4$, the spectral index is blue, $n > 1$, which is ruled out
by observation. The WMAP $2 \sigma$ limit $n = 0.96 \pm 0.026$
\cite{Komatsu:2008hk} corresponds to $s = [-3.814,-3.959]$. Since
the Hubble horizon is growing in comoving units, no gravitational
wave modes are produced.

Tachyacoustic models are not a fully convincing alternative to
inflation, since they solve only the horizon problem and not the
flatness problem, and inflation solves both at once. However,
inflation has initial conditions problems of its own, in
particular the fact that the initial inflationary ``patch'' must
be larger than a horizon size for inflation to start
\cite{Vachaspati:1998dy}. Furthermore it has been shown that
inflationary spacetimes are in general geodesically
past-incomplete \cite{Borde:2001nh}. The initial conditions for
tachyacoustic cosmology are quite different than those for
inflation due to the presence of a true ``Big Bang'' singularity
at zero time. However, in this limit, the sound speed is {\it
infinite} and the tachyacoustic solution approaches an instanton.
To see this, examine the form of the DBI field Lagrangian
(\ref{eq:DBIlagrangian}) near the $\tau = 0$ boundary of a
tachyacoustic spacetime. From Eq. (\ref{eq:DBIsoundspeed}), the
$c_S \rightarrow \infty$ limit corresponds to $\phi \rightarrow
\infty$ and  $f\left(\phi\right) \dot\phi^2 \rightarrow -\infty$,
so that
\begin{equation}
{\mathcal L} \rightarrow \frac{\dot\phi}{\sqrt{\left\vert f\right\vert}} - V\left(\phi\right),
\end{equation}
where
\begin{equation}
\dot\phi \equiv \sqrt{g^{\mu\nu} \partial_\mu \phi \partial_\nu \phi}.
\end{equation}
From Eq. (\ref{eq:potentials}), the asymptotic behavior of $V\left(\phi\right)$ and $f\left(\phi\right)$ are
\begin{eqnarray}
V\left(\phi\right) &\rightarrow& 3 M_P^2 H^2 \propto \phi^{-4 \epsilon / s},\cr
f\left(\phi\right) &\rightarrow& -\frac{1}{2 M_P^2 \epsilon} \frac{c_S}{H} \propto \phi^{2\left(1 + 2 \epsilon / s\right)}.
\end{eqnarray}
The scale-invariant limit $s = -2 \epsilon$ is especially interesting, since
\begin{equation}
\frac{1}{\sqrt{\left\vert f\right\vert}} \rightarrow \mu^2 = {\rm const.},
\end{equation}
and the Lagrangian takes the form
\begin{equation}
\label{eq:Lcuscuton}
{\mathcal L} \rightarrow \mu^2 \dot\phi - V\left(\phi\right),
\end{equation}
where $V\left(\phi\right) \propto \phi^2$. This can be identified
as exactly the ``cuscuton'' Lagrangian, suggested by Afshordi,
{\it et al.} as a candidate for Dark Energy
\cite{Mukhanov:2005bu,Afshordi:2006ad,Afshordi:2007yx}. Similarly, the $n = 0$
solution considered in Sec. \ref{sec:reconstruction} approaches a
cuscuton on the initial boundary surface.  The cuscuton is a
non-dynamical, instanton-like solution with infinite speed of
sound. Consider the action for the Lagrangian
(\ref{eq:Lcuscuton}),
\begin{eqnarray}
\label{eq:Scuscuton}
S_\phi &=& \int{d^4 x \sqrt{-g} \left[\mu^2 \dot\phi - V\left(\phi\right)\right]}\cr
&=& \mu^2 \int{dt \dot\phi} \int{d^3 x \sqrt{-g}} -  \int{d^4 x \sqrt{-g} V\left(\phi\right)}\cr
&=&  \mu^2 \int{d\phi \Sigma\left(\phi\right)} - \int{d^4 x \sqrt{-g} V\left(\phi\right)},
\end{eqnarray}
where $\Sigma\left(\phi\right)$ is the volume of a constant-$\phi$
hypersurface in the spacetime. The classical solutions to the
cuscuton action are constant mean curvature hypersurfaces,
analogous to soap bubbles \cite{Afshordi:2006ad}. It is
interesting to speculate that this property of the cuscuton action
may provide a self-consistent cosmological boundary condition, or
(even more speculatively) be useful as a solution to the
cosmological flatness problem. A full analysis, however, would
require inclusion of the gravitational action and solution in a
Wheeler-De Witt framework, or perhaps an embedding of the model
in string theory or an alternate gravity theory such as Horava-Lifshitz \cite{Piao:2009ax,Afshordi:2009tt}.
This is the subject of future work.

\appendix
\section{Stable Causality}
\label{sec:stablecau}

Since tachyacoustic cosmology deals with superluminal propagation of perturbations, it is important to address the issue of {\it{causality}} in this
model. Babichev {\it{et. al.}} \cite{Babichev:2007dw} have discussed the conditions that must be fulfilled by a general k-essence model with
superluminal propagation in order to avoid causal paradoxes ({\it{i.e.}}, the presence of {\it{closed causal curves}} - CCC). In this appendix we
outline the main ideas of this work and apply to our tachyacoustic model.

To begin with let us introduce some key definitions \cite{Wald:1984rg}. Let $g_{\mu\nu}$ be a metric with Lorentzian signature defined on a given manifold
${\cal{M}}$. Given a point $p\in {\cal{M}}$, let $t^{\mu}$ be a timelike vector at $p$; then, from this timelike vector we construct a second metric,
$\tilde{g}_{\mu\nu}$, related to the background metric $g_{\mu\nu}$ by
\begin{equation}
\label{stcau}
\tilde{g}_{\mu\nu}\equiv g_{\mu\nu}-t_{\mu}t_{\nu},
\end{equation}
which clearly has a Lorentzian signature. Then, the spacetime $\left({\cal{M}},g_{\mu\nu}\right)$ is said to be {\it{stably causal}} if there is a
continuous timelike vector field $t^{\mu}$ such that the spacetime $\left({\cal{M}},\tilde{g}_{\mu\nu}\right)$ possesses no closed timelike curves.
The following theorem (8.2.2. in \cite{Wald:1984rg}) establishes the necessary and sufficient conditions for a spacetime to be stably causal:
\\
\\
{\it{A spacetime}} $\left({\cal{M}},g_{\mu\nu}\right)$ {\it{stably causal if and only if there exists a differentiable function f on ${\cal{M}}$ such
that $\nabla^{\mu}f$ is a past directed timelike vector field}}.
\\
\\
We can apply this theorem to k-essence models as follows \cite{Babichev:2007dw}. First, we must find the analog of the induced metric (\ref{stcau}) for
the case of k-essence models, which can be obtained by means of the equation of motion for a scalar field described by a Lagrangian
${\cal{L}}\left(X,\phi\right)$,
\begin{equation}
\label{eom}
\tilde{G}^{\mu\nu}\nabla_{\mu}\nabla_{\nu}\phi+2X{\cal{L}}_{X\phi}-{\cal{L}}_{\phi}=0,
\end{equation}
where $\tilde{G}^{\mu\nu}$, called ``effective" or ``acoustic" metric is given by
\begin{equation}
\label{defeffmet}
\tilde{G}^{\mu\nu}\left(\phi,\nabla\phi\right)={\cal{L}}_{X}g^{\mu\nu}+{\cal{L}}_{XX}\nabla^{\mu}\phi\nabla^{\nu}\phi.
\end{equation}
It is convenient to use the metric \cite{Babichev:2007dw}
\begin{equation}
\label{confdefeffmet}
G^{\mu\nu}\equiv\frac{c_S}{{\cal{L}}_{X}^2}\tilde{G}^{\mu\nu}
\end{equation}
which is conformally equivalent to $\tilde{G}^{\mu\nu}$, and hence, defines the same causal structure. The inverse metric $G^{-1}_{\mu\nu}$ is given by
\begin{equation}
\label{invdefeffmet}
G^{-1}_{\mu\nu}\equiv\frac{{\cal{L}}_{X}}{c_S}\left[g_{\mu\nu}-c_S^2\frac{{\cal{L}}_{XX}}{{\cal{L}}_{X}}\nabla_{\mu}\phi\nabla_{\nu}\phi\right];
\end{equation}
notice that it has the same form of (\ref{stcau}), since
$\nabla^{\mu}\phi$ is timelike. Using this definition, we can now
apply the theorem stated above and check the stable causality of
k-essence models. Let $t$ be time coordinate with respect to the
background metric (which is everywhere future directed), which we
take to be FRW. Since $\nabla_{\mu}t\nabla_{\nu}t=1$, we have,
using (\ref{defeffmet}) and (\ref{confdefeffmet}),
\begin{equation}
\label{prstcau1}
G^{\mu\nu}\nabla_{\mu}t\nabla_{\nu}t=\frac{c_S}{{\cal{L}}_{X}}\left[1+\frac{{\cal{L}}_{XX}}{{\cal{L}}_{X}}\dot{\phi}^2\right];
\end{equation}
then, since for a {\it{homogeneous}} scalar field holds $\dot{\phi}^2=2X$, we have, from (\ref{defspeedofsound}) and (\ref{prstcau1}) that
\begin{equation}
\label{prstcau}
G^{\mu\nu}\nabla_{\mu}t\nabla_{\nu}t=\frac{1}{c_S{\cal{L}}_{X}}>0,
\end{equation}
\\
provided the Null Energy Condition (NEC) is satisfied, that is, ${\cal{L}}_{X}>0$. Therefore, $t$ plays a role of global time for {\it{both}} spacetimes
$\left({\cal{M}},g_{\mu\nu}\right)$ and $\left({\cal{M}},G^{-1}_{\mu\nu}\right)$, and then the conditions of the theorem are fulfilled. Then, there is
{\it{no}} CCC in superluminal k-essence models built from homogeneous scalar fields on a FRW background. Since this is exactly the case of the
models introduced in this paper, we conclude that there are no causal paradoxes in tachyacoustic cosmology.

\begin{acknowledgments}
This  research is supported  in part by the National Science Foundation under grant NSF-PHY-0757693. DB thanks the Brazilian agency CAPES for
financial support.

\end{acknowledgments}


\begin{thebibliography}{99}

\bibitem{Guth:1980zm}
   A.~H.~Guth,
   Phys.\ Rev.\ D {\bf 23}, 347 (1981).

\bibitem{Linde:1981mu}
   A.~D.~Linde,
   Phys.\ Lett.\ B {\bf 108}, 389 (1982).

\bibitem{Albrecht:1982wi}
   A.~Albrecht and P.~J.~Steinhardt,
   Phys.\ Rev.\ Lett.\  {\bf 48}, 1220 (1982).

\bibitem{Dunkley:2008ie}
  J.~Dunkley {\it et al.}  [WMAP Collaboration],
  Astrophys.\ J.\ Suppl.\  {\bf 180}, 306 (2009)
  [arXiv:0803.0586 [astro-ph]].

\bibitem{Komatsu:2008hk}
  E.~Komatsu {\it et al.}  [WMAP Collaboration],
  Astrophys.\ J.\ Suppl.\  {\bf 180}, 330 (2009)
  [arXiv:0803.0547 [astro-ph]].

\bibitem{Kinney:2008wy}
  W.~H.~Kinney, E.~W.~Kolb, A.~Melchiorri and A.~Riotto,
  Phys.\ Rev.\  D {\bf 78}, 087302 (2008)
  [arXiv:0805.2966 [astro-ph]].

\bibitem{Gratton:2003pe}
  S.~Gratton, J.~Khoury, P.~J.~Steinhardt and N.~Turok,
  Phys.\ Rev.\  D {\bf 69}, 103505 (2004)
  [arXiv:astro-ph/0301395].

\bibitem{Albrecht:1998ir}
  A.~J.~Albrecht and J.~Magueijo,
  Phys.\ Rev.\  D {\bf 59}, 043516 (1999)
  [arXiv:astro-ph/9811018].

\bibitem{Dutta:2005gt}
  S.~Dutta and T.~Vachaspati,
  Phys.\ Rev.\  D {\bf 71}, 083507 (2005)
  [arXiv:astro-ph/0501396].

\bibitem{Dutta:2005if}
  S.~Dutta,
  Phys.\ Rev.\  D {\bf 73}, 063524 (2006)
  [arXiv:astro-ph/0511120].

\bibitem{Brandenberger:2009jq}
  R.~H.~Brandenberger,
  arXiv:0902.4731 [hep-th].

\bibitem{Starkman:2000dy}
  G.~D.~Starkman, D.~Stojkovic and M.~Trodden,
  Phys.\ Rev.\  D {\bf 63}, 103511 (2001)
  [arXiv:hep-th/0012226].

\bibitem{Starkman:2001xu}
  G.~D.~Starkman, D.~Stojkovic and M.~Trodden,
  Phys.\ Rev.\ Lett.\  {\bf 87}, 231303 (2001)
  [arXiv:hep-th/0106143].

\bibitem{Khoury:2008wj}
  J.~Khoury and F.~Piazza,
  arXiv:0811.3633 [hep-th].

\bibitem{ArmendarizPicon:2006if}
   C.~Armendariz-Picon,
   JCAP {\bf 0610}, 010 (2006)
   [arXiv:astro-ph/0606168].

\bibitem{Piao:2006ja}
  Y.~S.~Piao,
  Phys.\ Rev.\  D {\bf 75}, 063517 (2007)
  [arXiv:gr-qc/0609071].

\bibitem{Magueijo:2008pm}
  J.~Magueijo,
  Phys.\ Rev.\ Lett.\  {\bf 100}, 231302 (2008)
  [arXiv:0803.0859 [astro-ph]].

\bibitem{Magueijo:2008sx}
  J.~Magueijo,
  Phys.\ Rev.\  D {\bf 79}, 043525 (2009)
  [arXiv:0807.1689 [gr-qc]].

\bibitem{Piao:2008ip}
  Y.~S.~Piao,
  arXiv:0807.3226 [gr-qc].

\bibitem{Kinney:2002qn}
  W.~H.~Kinney,
  Phys.\ Rev.\  D {\bf 66}, 083508 (2002)
  [arXiv:astro-ph/0206032].

\bibitem{Bean:2008ga}
  R.~Bean, D.~J.~H.~Chung and G.~Geshnizjani,
  Phys.\ Rev.\  D {\bf 78}, 023517 (2008)
  [arXiv:0801.0742 [astro-ph]].

\bibitem{ArmendarizPicon:1999rj}
  C.~Armendariz-Picon, T.~Damour and V.~F.~Mukhanov,
  Phys.\ Lett.\  B {\bf 458}, 209 (1999)
  [arXiv:hep-th/9904075].

\bibitem{Silverstein:2003hf}
  E.~Silverstein and D.~Tong,
  Phys.\ Rev.\  D {\bf 70}, 103505 (2004)
  [arXiv:hep-th/0310221].

\bibitem{Garriga:1999vw}
  J.~Garriga and V.~F.~Mukhanov,
  Phys.\ Lett.\  B {\bf 458}, 219 (1999)
  [arXiv:hep-th/9904176].

\bibitem{Babichev:2006vx}
  E.~Babichev, V.~F.~Mukhanov and A.~Vikman,
  JHEP {\bf 0609}, 061 (2006)
  [arXiv:hep-th/0604075].

\bibitem{Babichev:2007dw}
  E.~Babichev, V.~Mukhanov and A.~Vikman,
  JHEP {\bf 0802}, 101 (2008)
  [arXiv:0708.0561 [hep-th]].

\bibitem{Chimento:2007es}
  L.~P.~Chimento and R.~Lazkoz,
  Gen.\ Rel.\ Grav.\  {\bf 40}, 2543 (2008)
  [arXiv:0711.0712 [hep-th]].

\bibitem{Kinney:2007ag}
  W.~H.~Kinney and K.~Tzirakis,
  Phys.\ Rev.\  D {\bf 77}, 103517 (2008)
  [arXiv:0712.2043 [astro-ph]].

\bibitem{Klebanov:2000hb}
  I.~R.~Klebanov and M.~J.~Strassler,
  JHEP {\bf 0008}, 052 (2000)
  [arXiv:hep-th/0007191].

\bibitem{Muslimov:1990be}
  A.~G.~Muslimov,
  Class.\ Quant.\ Grav.\  {\bf 7}, 231 (1990).

\bibitem{Salopek:1990jq}
  D.~S.~Salopek and J.~R.~Bond,
  Phys.\ Rev.\  D {\bf 42}, 3936 (1990).

\bibitem{Lidsey:1995np}
  J.~E.~Lidsey, A.~R.~Liddle, E.~W.~Kolb, E.~J.~Copeland, T.~Barreiro and M.~Abney,
  Rev.\ Mod.\ Phys.\  {\bf 69}, 373 (1997)
  [arXiv:astro-ph/9508078].

\bibitem{Kinney:1997ne}
  W.~H.~Kinney,
  Phys.\ Rev.\  D {\bf 56}, 2002 (1997)
  [arXiv:hep-ph/9702427].

\bibitem{Spalinski:2007kt}
  M.~Spalinski,
  JCAP {\bf 0704}, 018 (2007)
  [arXiv:hep-th/0702118].

\bibitem{Abramowitz:1972}
   M.~Abramowitz and I.~A.~Stegun, {\it{Handbook of Mathematical Functions}}, New York: Dover, 1972.


\bibitem{Afshordi:2006ad}
  N.~Afshordi, D.~J.~H.~Chung and G.~Geshnizjani,
  Phys.\ Rev.\  D {\bf 75}, 083513 (2007)
  [arXiv:hep-th/0609150].

\bibitem{Vachaspati:1998dy}
  T.~Vachaspati and M.~Trodden,
  Phys.\ Rev.\  D {\bf 61}, 023502 (1999)
  [arXiv:gr-qc/9811037].

\bibitem{Borde:2001nh}
  A.~Borde, A.~H.~Guth and A.~Vilenkin,
  Phys.\ Rev.\ Lett.\  {\bf 90}, 151301 (2003)
  [arXiv:gr-qc/0110012].

\bibitem{Lidsey:2007gq}
  J.~E.~Lidsey and I.~Huston,
  JCAP {\bf 0707}, 002 (2007)
  [arXiv:0705.0240 [hep-th]].

\bibitem{Afshordi:2007yx}
  N.~Afshordi, D.~J.~H.~Chung, M.~Doran and G.~Geshnizjani,
  Phys.\ Rev.\  D {\bf 75}, 123509 (2007)
  [arXiv:astro-ph/0702002].

\bibitem{Mukhanov:2005bu}
  V.~F.~Mukhanov and A.~Vikman,
  JCAP {\bf 0602}, 004 (2006)
  [arXiv:astro-ph/0512066].

\bibitem{Piao:2009ax}
  Y.~S.~Piao,
  arXiv:0904.4117 [hep-th].

\bibitem{Afshordi:2009tt}
  N.~Afshordi,
  arXiv:0907.5201 [hep-th].

\bibitem{Wald:1984rg}
  R.~M.~Wald,
  ``General Relativity,''
  Chicago, USA: Univ. Pr. ( 1984)



\end{thebibliography}
\end{document}